# Alternative method to find orbits in chaotic systems

Kai T. Hansen[*]

*Fakultät für Physik, Universität Freiburg*
*Hermann-Herder-Strasse 3, D-79104 Freiburg, Germany*
*e-mail: k.t.hansen@fys.uio.no*

We present here a new method which applies well ordered symbolic dynamics to find unstable periodic and non-periodic orbits in a chaotic system. The method is simple and efficient and has been successfully applied to a number of different systems such as the Hénon map, disk billiards, stadium billiard, wedge billiard, diamagnetic Kepler problem, colinear Helium atom and systems with attracting potentials. The method seems to be better than earlier applied methods.

To analyze chaotic a system is it often necessary to find the periodic orbits in a system. This is especially important when applying semi-classical methods to investigate quantum mechanical systems. The periodic orbits (cycles) yields a skeleton for the dynamics [1,2]. We restrict ourself here to flows with two-dimensional Poincaré maps and to maps of a two-dimensional plane.

The problem of numerically finding the existing cycles is not trivial, and different methods have been applied. The problem can be formulated as finding the zero points of complicated two-dimensional functions. This can be solved by using the Newton-Raphson method, which converges very fast when the initial point is close to the correct solution. One problem with this method is that good initial points are very hard to find for long periodic orbits. If one does not find a cycle it is difficult to know if the cycle does not exist or if one did not have a sufficiently good initial guess.

For some systems there exist special methods. In a dispersing billiard system it is possible to find all orbits using a "rubber band" method. This is simply done by choosing randomly one point on each boundary according to a description of which boundaries the particle should bounce off, and then vary these boundary points until the length of the line connecting them reach a minimum. This is a "safe" method as it always converges to a solution which may be the cycle or a non-physical solution in the case where the cycle does not exist. This method however converges very slowly for long orbits and is therefore not ideal. A similar method has been used by Biham and Kvale for the stadium billiard [3].

In the case of the Hénon map there exists a method proposed by Biham and Wenzel [4] which converges to a cycle with a given symbolic dynamics description. This method has a rather slow convergence and for some parameter values the method will not converge [5].

More complicated methods have also been applied to find cycles in Hamiltonian systems [6].

The method proposed here works as follows: Let $\delta$ and $\gamma$ be defined as the well ordered symbolic values for the past and future respectively (discussed below for specific examples). Assume we want to find the cycle with the symbolic string $\overline{S} = \overline{s_1 s_2 \cdots s_n} = \cdots s_1 s_2 \cdots s_n s_1 s_2 \cdots s_n \cdots$. We then make the following calculations

- Find the symbolic values $(\delta_c, \gamma_c)$ for the cycle $\overline{S}$.

- Calculate $(\delta, \gamma)$ for a chosen starting point $(x, y)$ in the Poincaré plane.

- Estimate the direction of the $\gamma$- and $\delta$-axis in the Poincaré plane close to the point $(x, y)$.

- Move the starting point in the Poincaré plane by comparing $(\delta, \gamma)$ for this point with the $(\delta, \gamma)$ values for the cycle.

Implemented as a binary search this procedure converges fast to the correct starting point $(x, y)$ for the periodic orbit.

The $\gamma$-axis should be chosen parallel to the unstable manifolds and the $\delta$-axis should be parallel to the stable manifold at the point $(x, y)$, but even rather large errors here only give a slow convergence and the method is very robust. It is usually enough to assume that a fixed direction of the $\gamma$- and $\delta$-axis is valid in the whole Poincaré plane. In the neighborhood of a homoclinic tangency this has to be done more carefully. If the system is a scattering system the method converges equally fast as for a bounded system using a proper definition of $\gamma$ and $\delta$ for an escaping orbit. The binary search is done by using parallelograms in the Poincaré plane defined by two corners and compare the $\delta$ and $\gamma$ values of the midpoint of the parallelogram with the correct values. When new untested points are chosen as corner in a new and smaller parallelogram, the symbolic values of these points have to be tested since the manifolds are not exactly parallel to the $\gamma$- and $\delta$-axis. This makes the method slightly slower than a straight forward one-dimensional binary search.

If we want to find a non-periodic orbit with some particular symbolic description $S$, for example a homoclinic or a heteroclinic orbit, this is done by using $(\delta, \gamma)$ for this orbit. For a generic orbit which does not emerge from nor converge to a periodic orbit, $\gamma$ and $\delta$ are not rational numbers.

As a simple application of the method we describe how it is implemented for the Hénon map [7]. The Hénon map is defined as

$$x_{t+1} = 1 - ax_t^2 + bx_{t-1} \tag{1}$$

The Poincaré plane is here simply the phase space plane $(x_t, x_{t+1})$. For the parameter $a$ sufficiently large this map yields a complete Smale horseshoe [8]. If $|b| = 1$ the map is area preserving. In the case of a Smale horseshoe the symbolic dynamics description for a starting point $(x_0, x_1)$ is given by the symbol string $S = \cdots s_{-2} s_{-1} s_0 \cdot s_1 s_2 s_3 \cdots$ with $s_t$ defined as

$$s_t = \begin{cases} 0 & \text{if} \quad x_t < 0 \\ 1 & \text{if} \quad x_t > 0 \end{cases} \tag{2}$$

If the map is not a complete repellor we may use the partition defined by primary turning points as suggested by Grassberger and Kantz [9]. Well ordered symbols and the symbol plane $(\delta, \gamma)$ for this map was introduced by Cvitanović and coworkers [10] to describe all admissible orbits by determining a pruning front in this plane. The definition of $\delta$ depends on the sign of the parameter $b$ because for $b < 0$ the map only folds the phase space once while for $b > 0$ the map both folds the phase space once and then reflects it yielding a different ordering of the unstable foliation. Both for $b < 0$ and for $b > 0$ is $\gamma$ defined as

$$w_{t+1} = \begin{cases} w_t & \text{if } s_{t+1} = 0 \\ 1 - w_t & \text{if } s_{t+1} = 1 \end{cases}, \quad w_1 = s_1$$

$$\gamma(S) = 0.w_1 w_2 w_3 \ldots = \sum_{t=1}^{\infty} w_t/2^t. \tag{3}$$

The symbolic past $\delta$ is defined for $b < 0$ as

$$w_{t-1} = \begin{cases} w_t & \text{if } s_{t-1} = 0 \\ 1 - w_t & \text{if } s_{t-1} = 1 \end{cases}, \quad w_0 = s_0$$

$$\delta(S) = 0.w_0 w_{-1} w_{-2} \ldots = \sum_{t=1}^{\infty} w_{1-t}/2^t. \tag{4}$$

and for $b > 0$ as

$$w_{t-1} = \begin{cases} 1 - w_t & \text{if } s_{t-1} = 0 \\ w_t & \text{if } s_{t-1} = 1 \end{cases}, \quad w_0 = s_0$$

$$\delta(S) = 0.w_0 w_{-1} w_{-2} \ldots = \sum_{t=1}^{\infty} w_{1-t}/2^t. \tag{5}$$

For a period $n$ orbit $\overline{s_1 s_2 \cdots s_n}$ we have

$$\gamma = \left( \sum_{t=1}^{2n} w_t/2^t \right) 2^{2n}/(2^{2n} - 1)$$

and

$$\delta = \left( \sum_{t=1}^{2n} w_{1-t}/2^t \right) 2^{2n}/(2^{2n} - 1).$$

The simplest assumption is to choose the $\delta$-axis parallel to the $x_t$-axis and the $\gamma$-axis parallel to the $x_{t+1}$-axis in the whole $(x_t, x_{t+1})$ plane. In Fig.1 we find that this yields approximately the $\delta$-axis parallel to the stable manifolds and the $\gamma$-axis parallel to the unstable manifolds close to all crossing points between the stable and unstable manifolds. The parameter values are for this example $a = 3$ and $b = 0.3$ which gives a complete Smale horseshoe repellor.

We choose a large rectangle defined by a lower-left corner $(x_{\min}, x'_{\min})$ and a upper-right corner $(x_{\max}, x'_{\max})$. We then calculate the values $\delta$ and $\gamma$ for the middle point $((x_{\min} + x_{\max})/2, (x'_{\min} + x'_{\max})/2)$ and choose the new rectangle to be

$$
\begin{array}{ll}
\left(\frac{x_{\min}+x_{\max}}{2}, \frac{x'_{\max}+x'_{\max}}{2}\right), (x_{\max}, x'_{\max}) & \text{if} \quad \delta < \delta_c, \gamma < \gamma_c \\
(x_{\min}, x'_{\min}), \left(\frac{x_{\min}+x_{\max}}{2}, \frac{x'_{\min}+x'_{\max}}{2}\right) & \text{if} \quad \delta > \delta_c, \gamma > \gamma_c \\
\left(\frac{x_{\min}+x_{\max}}{2}, x'_{\min}\right), \left(x_{\max}, \frac{x'_{\max}+x'_{\max}}{2}\right) & \text{if} \quad \delta < \delta_c, \gamma > \gamma_c \\
\left(x_{\min}, \frac{x'_{\max}+x'_{\max}}{2}\right), \left(\frac{x_{\min}+x_{\max}}{2}, x'_{\max}\right) & \text{if} \quad \delta > \delta_c, \gamma < \gamma_c
\end{array}
\qquad (6)
$$

In case of the two last choices we do have to check the new corner points and move the lower-left point slightly down if the $\gamma$ value for this point is larger than $\gamma_c$ and move it slightly to the left is the $\delta$ value is larger than $\delta_c$, and correspondingly for the upper-right corner.

If a trajectory escapes we let the last symbol $w_t$ be 0 or 2 depending on the number of preceding symbols with $s = 1$ and let $\gamma$ and $\delta$ be given by a finite sum. This gives the symbolic value for the folds at the border the escaping region. Implemented on a computer is it faster to just compare symbols $w_t$ than to calculate the values $\gamma$ and $\delta$.

Instead of the binary search one could use a secant method which converges faster if one has a nice smooth function. The points $\delta$ and $\gamma$ vary as a monotone devils staircase type of function with $x$ and $y$ (continous if the repellor is complete), and numerics yields that the simplest binary seach (bisection method) is faster for these fractal functions. Nummerics indicate that a secant method could converge faster if the devils staircase function only has short horizontal intervals, but not much is known of the convergence properties of such functions.

A straight forward implementation of this algorithm for the repellor in Fig.1 gives the starting points with an error less than $10^{-30}$ for all 4720 cycles up to length 15 in around 5 CPU-minutes on a normal work-station. We have also examined the method for the parameter values $a = 1.4$, $b = 0.3$ which yields a strange attractor, and for other parameter values.

This algorithm is simple to program and we can use this method for many other systems. For the dispersing billiards consisting of 3-disks and 4-disks the algorithm for the symbolic values $(\delta, \gamma)$ is given in ref. [11]. Assuming the $\delta$- and $\gamma$-axis are along lines $\pm mx$ in the Poincaré plane with $(x, \phi)$ the posisiton and angle of a bounce the method converges to the correct orbit both for the complete repelling system, for the pruned repellor and for the overlapping disk bounded system. The number $m$ depends on the parameters but can usually be chosen around 1.0. In this problem it gives no problems if there are pruned orbits, because all manifolds are sharply broken or ends in a point and there are no homoclinic tangencies.

The chaotic Wedge billiard [12,5] also yields a similar foliation of manifolds. The values of $\gamma$ and $\delta$ are given by eq. (3) and (4) and the method works without problems.

The Bunimovich stadium billiard has a more complicated symbolic dynamics. The well ordered symbolic values $\gamma$ and $\delta$ are defined in ref. [13]. Assuming $\gamma$- and $\delta$-axis fixed at $\pm mx$ as for the disk systems, gives a working method to find the periodic orbits.

A physically interesting system is the chaotic colinear Helium atom [14]. A Poincaré map of this system using the point $(r_1, p_1)$ each time $r_2 = 0$ (with $r_1, p_1, r_2, p_2$ position and momentom of the two elektrons, see definitions in ref. [14]) yields a plane with manifolds ordered as a binary Cantor set and well ordered symbols are obtained from eq. (3) and (4). This Poincaré plane is unbounded since there exist long cycles with arbitrary large $r_1$ values, but this is not any problem when we only want to find orbits with a finite length. This system is a complete repelling binary Cantor set and we do not get any problems from homoclinic tangencies.

The diamagnetic Kepler problem [15] (a classical Hydrogen atom in a homogeneous magnetic field) is a completely chaotic trinary repellor for sufficiently large scaled energies [16]. The orbits can be described by the same symbolic alphabet as the 4-disk system [17] and symbolic values are defined as for this system. For the complete repellor the symbolic dynamics is easy to determine and the method of finding orbits converges nicely using a similar Poincaré map as for the colinear Helium. By numerically integrating this flow we can find closing cycles in a few CPU-secunds and starting points with 15 correct digits in a few CPU-minutes. We have calculated cycles as long as 1500 symbols, and we have applied this method in successfully calculating quantum mechanical resonances for Hydrogen in a magnetic field. For the pruned repellor case and for the bounded system one can describe the admissible orbits in this system using a pruning front description [18]. For these parameter values the symbols have to be defined in a more careful way and for some orbits one has to choose $\gamma$- and $\delta$-axix with care.

A different system where the presented method works is a Hamiltonian flow in two dimensions with three attracting Gaussian potentials [19]. For different parameter values the dynamics can yield a binary Cantor set repellor or it may be a 6-ary, 8-ary, … Cantor repellor and a symbolic dynamics can be given to the system. In the binary case we define a "inner bent orbit" $s = 1$ and a "outer bent orbit" $s = 0$. Then eq. (3) and (4) yields the symbolic values and the method converges. A similar system with two Lenard-Jones potentials yields for some parameters a 5 Cantor set repellor and using a 5-letter well ordered alphabet we find periodic orbits in this system. The structure and symbolic dynamics of these systems will be presented in a later article.

We expect this method to be applicable to all systems which can be described with a well ordered symbolic alphabet. If one wants to find an extremely long orbit, or an orbit with very high accuracy, then is it possible to find all points $(x_t, y_t)$ with $t \in \{0, 1, 2 \ldots, (n-1)\}$ by shifting the symbol string $t$-times and repeat the calculation finding $(x_t, y_t)$ by the method above. When integrating along such an orbit one adjusts the position at each crossing of the Poincaré plane. We have however never found this necessary in any application.

The author thanks for useful discussions Jan-Michael Rost and Predrag Cvitanović, the leading cyclist. The author is grateful to the the Alexander von Humboldt foundation for financial support.

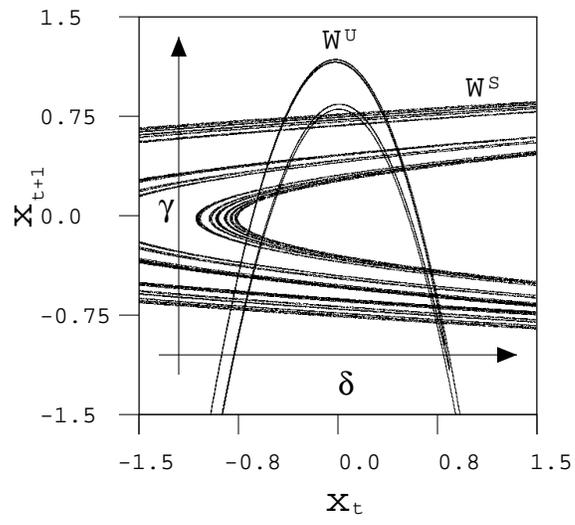

FIG. 1. Stable and unstable manifolds ($W^s$ and $W^u$) of the Hénon map $a = 3$ and $b = 0.3$ and the estimated direction of the $\delta$- and $\gamma$-axis.